\documentclass{aa}  
\pdfoutput=1
\usepackage{graphicx}
\usepackage{color}
\usepackage{amsmath}
\usepackage{natbib}
\usepackage{txfonts}
\usepackage{hyperref}
\hypersetup{breaklinks=true, colorlinks=true, citecolor=blue, linkcolor=blue,
urlcolor=blue}
\begin{document} 




 \title{What can the 2008/10 broadband flare of PKS\,1502+106 tell us?}

   \subtitle{Nuclear opacity, magnetic fields, and the
location of $\gamma$ rays}

   \author{V. Karamanavis\inst{1}\thanks{To whom correspondence should be
addressed.}
          \and
          L. Fuhrmann\inst{1}
          \and
          E. Angelakis\inst{1}
          \and
          I. Nestoras\inst{1}
          \and
          I. Myserlis\inst{1}
          \and
          T. P. Krichbaum\inst{1}
          \and
          J. A. Zensus\inst{1}
          \and
          \\H. Ungerechts\inst{2}
          \and
          A. Sievers\inst{2}
          \and
          M. A. Gurwell\inst{3}
          }

   \institute{Max-Planck-Institut f\"{u}r Radioastronomie, Auf dem H\"{u}gel
69, D-53121 Bonn, Germany\\
              \email{\href{mailto:vkaramanavis@mpifr.de}{vkaramanavis@mpifr.de}}
              \and
        Instituto de Radio Astronom\'{i}a Milim\'{e}trica, Avenida Divina
Pastora 7, Local 20, E-18012, Granada, Spain
              \and
        Harvard-Smithsonian Center for Astrophysics, Cambridge, MA 02138 USA
         }

   \date{Received ---, ---; accepted 11 March 2016}


  \abstract
  {The origin of blazar variability, seen from radio up to $\gamma$ rays, is
still a heavily debated matter and broadband  flares offer a unique
testbed towards a better understanding of these extreme objects. Such an
energetic outburst was detected by \textit{Fermi}/LAT in 2008 from the blazar
PKS\,1502+106. The outburst was observed from $\gamma$ rays down to
radio frequencies.}
  {Through the delay between flare maxima at different radio frequencies, we
study the frequency-dependent position of the unit-opacity surface and infer its
absolute position with respect to the jet base. This nuclear opacity profile
enables the magnetic field tomography of the jet. We also localize the
$\gamma$-ray emission region and explore the mechanism producing the flare.}
  {The radio flare of PKS\,1502+106 is studied through single-dish
flux density measurements at 12 frequencies in the range 2.64
to 226.5\,GHz. To quantify it, we employ both a Gaussian process regression and
a discrete cross-correlation function analysis.}
  {We find that the light curve parameters (flare amplitude and cross-band
delays) show a power-law dependence on frequency. Delays decrease with
frequency, and the flare amplitudes increase up to about 43\,GHz and then decay.
This behavior is consistent with a shock propagating downstream the jet. The
self-absorbed radio cores are located between ${\sim}10$ and 4\,pc from
the jet base and their magnetic field strengths range between 14 and
176\,mG, at the frequencies 2.64 to 86.24\,GHz. Finally, the $\gamma$-ray active
region is located at $(1.9 \pm 1.1)$\,pc away from the jet base.} 
  {}


   \maketitle
%

\section{Introduction}\label{sect:intro}

Blazars comprise the radio-loud minority of active galactic nuclei (AGN) with
energetic jets seen at small viewing angles. Observationally, they exhibit
strong variability across the whole electromagnetic spectrum. However, its
origin and extreme characteristics (broadband flux density outbursts with down
to minute-long time scales) are still a matter of intense debate. Moreover,
uncertainty still enshrouds the exact location of its high-energy component,
observed at MeV--TeV energies.

A multitude of mechanisms has been proposed in order to explain this intense
activity. Among those are: shocks generated and propagating along the
relativistic jet \citep[e.g.][]{1985ApJ...298..114M, 1992A&A...254...71V},
colliding relativistic plasma shells \citep[][]{2001MNRAS.325.1559S}, precession
of the jet \citep[e.g.][]{2003ApJ...584..135L}, helical motion in it
\citep[e.g][]{1992A&A...255...59C}, or Kelvin--Helmholtz instabilities
developing in the outflow \citep[e.g.][]{2001Sci...294..128L}.
Testing the validity of such models translates into quantitatively
contrasting the observational signatures of the mechanisms they invoke, with
observations. Multi-frequency flux density monitoring can shed light onto
blazar physics by accessing time and thus spatial scales unreachable even by
interferometric imaging. Dense, long-term, multi-frequency monitoring offers a
unique opportunity for putting blazar variability models to the test and
comparing with the predictions of theoretical frameworks that attempt to explain
the emission and overall observed characteristics of those sources.

Time delays between flare maxima are often seen, and usually attributed to
opacity effects. Flare onsets and maxima appear first at higher
frequencies and successively progress towards the lower end of the observing
frequency range \citep[e.g.][]{1992A&A...254...71V}. This is usually connected
to the motion of disturbances whose optical depth decreases, while moving
downstream the jet \citep[e.g.][]{1985ApJ...298..114M} and expanding
adiabatically \citep[][]{1966Natur.211.1131V}. Consequently, lower frequency
(hence energy) photons are able to escape the synchrotron-emitting region.

PKS\,1502+106 (OR\,103, S3\,1502+10) at a redshift of $z = 1.8385$, $\rm{D_{L}}
= 14176.8$ Mpc \citep{2008ApJS..175..297A}, is a powerful flat spectrum radio
quasar (FSRQ) harboring a supermassive black hole of ${\sim}
10^{9}$M$_{\odot}$ \citep[][and references therein]{2010ApJ...710..810A}. The
focal point here, is the intense 2008/2010 outburst seen in 
PKS\,1502+106 \citep{2008ATel.1650....1C}. This pronounced broadband flare,
observed from radio up to $\gamma$-ray energies, triggered the first
\textit{Fermi}-GST multi-frequency campaign covering the electromagnetic
spectrum in its entirety, both with filled-aperture instruments and 
interferometric arrays \citep[][see also
\cite{2015PhDT.......232K} for first results]{2016A&A...586A..60K}.

The scope here is to parameterize the observed outburst and extract its
relevant light curve parameters such as, the flare amplitude at each frequency
and the cross-band delays. First, using the observed cross-band and
frequency-dependent time lags we estimate the core shift -- i.e. the
frequency-dependent position of the core -- with an approach other than via
traditional multi-frequency very-long-baseline interferometry (VLBI)
measurements. With the time-lag core shift at hand, an opacity profile of the
source is obtained and, under the assumption of equipartition, the magnetic
field strength along the synchrotron-self-absorbed jet. Furthermore, the
distance of each core to the vertex of the jet, in combination with the observed
delay between radio and $\gamma$ rays, allow for decisively constraining the
location of the high-energy emission. 


The paper is structured as follows. In Sect. \ref{sect:obsRed} we  present
the data set and the time series analysis techniques used. In Sect.
\ref{sect:results} our results are presented. Sect. \ref{sect:Disc} discusses
and Sect. \ref{sect:Conc} concludes our findings.
In the paper, we adopt $S \propto \nu^{+ \alpha}$, where $S$ is the
radio flux density, $\nu$ the observing frequency, and $\alpha$ the spectral
index, along with the following cosmological parameters $H_{0} = 71\,{\rm
km\,s}^{-1}\,{\rm Mpc}^{-1}$, $\Omega_{\rm m} = 0.27$, and $\Omega_{\Lambda} =
0.73$.

%
\section{Observations and data analysis}\label{sect:obsRed}

Data used in this work have been collected within the framework of the
F-GAMMA\footnote{\url{www.mpifr-bonn.mpg.de/div/vlbi/fgamma/fgamma.html}
} program \citep{2007AIPC..921..249F,2010arXiv1006.5610A}, which includes
observations with the Effelsberg 100-m telescope (EB) in eight frequency bands
from 2.64 to 43.05\,GHz, and with the IRAM 30-m telescope (PV) at 86.24 and
142.33\,GHz.
Monthly observations at EB and PV were performed in a quasi-simultaneous manner
with a typical separation of days ensuring maximum spectral coherency. A
detailed description of the observing setup and data reduction is provided
elsewhere \citep{2014MNRAS.441.1899F, 2015A&A...575A..55A, 2015Nestoras}.
We also employ data from the OVRO 40-m blazar monitoring
program\footnote{\url{www.astro.caltech.edu/ovroblazars}} at 15\,GHz
\citep{2011ApJS..194...29R} and at 226.5\,GHz from the SMA observer center data
base \citep{2007ASPC..375..234G}.


\subsection{Gaussian process regression}

In order to extract the light curve parameters, such as the flare amplitude and
time scale at each frequency, we employ a Gaussian process (GP) regression
scheme \cite[][]{Rasmussen:2005:GPM:1162254}.
This is a non-parametric approach to the generalized regression and prediction
problem, widely used in machine learning applications and lately exploited in
astronomy \citep[e.g.][]{2012MNRAS.419.2683G, 2012MNRAS.419.3147A}. While
traditional fitting assumes a specific functional form $f$ for the model, a
second way to {\it learn} $f$ from the data is by assigning a prior probability
to all
functions one considers more likely, for example on the grounds of their
smoothness; i.e. infinite differentiability. A Gaussian process is a probability
distribution over functions. It constitutes the generalization of the Gaussian
distribution of random variables or vectors, into the space of functions
\citep[e.g.][]{Rasmussen:2005:GPM:1162254, 10.2307/j.ctt4cgbdj}.

In the case of AGN light curve fitting, the problem has the form of
one-dimensional regression with the observables being flux density levels at
certain times. For instance, given a set of three observations, or training data
points, $\mathcal{D} = \{ (x_{1},y_{1}),(x_{2},y_{2}),(x_{3},y_{3}) \}$, our
problem reduces to finding the functions, drawn from the prior distribution,
that pass through all these training points.
This distribution of functions is the posterior distribution and its
mean, being the expected value, can be considered as the ``best-fit function''.
A Gaussian process is defined by its mean, $\mu$, and covariance, $k$.
Without loss of generality one can always assume that $\mu=0$ since the data 
can always be shifted to accommodate for this assumption.

In the context of Gaussian processes the covariance function,
covariance kernel, or simply kernel, ${\rm Cov}[f(x_{i}),
f(x_{j})] = k(x_{i}, x_{j})$ is used to define the covariance between any two
function values at points $x_{i}$ and $x_{j}$; i.e. the similarity between data
points. It is chosen on the basis of the prior beliefs for the function to be
learned. Essentially, the kernel defines the class of functions
that are likely to appear in the prior distribution which in turn determine the
kind of structure that the specific GP is able to model correctly. Kernels have
their own set of parameters, called hyperparameters, since they define
the properties of the prior distribution over functions, instead of the
functions themselves. A very useful property of kernels is that addition and
multiplication between two or more of them, also produce valid covariance
kernels. As such, there is always the option of constructing a covariance kernel
that fits the characteristics of the modelling problem\footnote{e.g.
\url{www.people.seas.harvard.edu/~dduvenaud/cookbook/}}. Here we employ the
squared exponential kernel
\begin{equation}
 k(x_{i}, x_{j}) = \sigma^{2} \exp \left[  \dfrac{-(x_{i}-x_{j})^{2}}{2l^{2}} 
\right],
\label{eq:SqExpKernel}
\end{equation}
where the hyperparameters are
\[
\begin{array}{lp{0.8\linewidth}}
  l          & the characteristic length scale, denoting the distance in the
$x$ dimension after which the function changes significantly, and \\
  \sigma^{2} & the variance, mapping the mean distance of the
function from its mean. Here, it serves only as a scaling factor. \\
\end{array}
\]
This is a widely used kernel, owing to its generality
and smoothness. The latter is the only assumption when using it, justified by
the fact that most time series arising from physical processes have no reason
not to be smooth. In the case of data with uncertainty, one only has to add it
to the noiseless covariance kernel. In the general case, that is
$\boldsymbol{k}_{n} (x_{i}, x_{j}) = \boldsymbol{k} (x_{i}, x_{j}) + \sigma^{2}
\boldsymbol{I}$, where $\boldsymbol{I}$ is the identity matrix
\citep[e.g.][]{GPpaper}.

Selecting the best set of hyperparameters using the data themselves is referred
to as training the GP or, more generally, Bayesian model
selection. In essence, one wants to update the prior knowledge in light of a
training data set.
One way of doing so is by maximizing the marginal likelihood
\citep{NIPS1995}. Letting $\boldsymbol{\theta}$ denote the vector of
hyperparameters, in the case of the squared exponential kernel
(Eq. \ref{eq:SqExpKernel}) we have
\begin{equation}
 \boldsymbol{\theta} = \{ l, \sigma^{2} \}.
\end{equation}
Then, the probability (or evidence) of the training data, $\boldsymbol{y}$,
given the hyperparameters vector $\boldsymbol{\theta}$, is 
\begin{equation}
 p(\boldsymbol{y} \,|\, \boldsymbol{x}, \boldsymbol{\theta} )
\end{equation}
and the log marginal likelihood is given by
\begin{equation}
\mathcal{L} = \log  p(\boldsymbol{y} \,|\, \boldsymbol{x}, \boldsymbol{\theta} )
=
 - \dfrac{1}{2} \boldsymbol{y}^{T} \boldsymbol{k}_{n}^{-1} \boldsymbol{y}
 - \dfrac{1}{2} \log |\boldsymbol{k}_{n}|
 - \dfrac{n}{2} \log 2 \pi.
\end{equation}
In the general case of a hyperparameter vector $\boldsymbol{\theta} = \{ 
\theta_{j} \,|\, j = 1, ..., n \} $, the derivatives of the log marginal
likelihood with respect to each $\theta_{j}$ are
\begin{equation}
 \dfrac{\partial \mathcal{L}}{\partial \theta_{j}} = 
\dfrac{1}{2} \boldsymbol{y}^{T} \dfrac{\partial
\boldsymbol{k}_{n}}{\partial \theta_{j}} \boldsymbol{k}_{n}^{-1}
\boldsymbol{y}
- \dfrac{1}{2} ~\mbox{Trace} \left( \boldsymbol{k}_{n}^{-1} 
\dfrac{\partial \boldsymbol{k}_{n}}{\partial \theta_{j}} \right).
\label{eq:gradient}
\end{equation}
Equation \ref{eq:gradient} can be used with any numerical gradient optimization
algorithm in order to maximize the log marginal likelihood and return the set of
best hyperparameters for the problem at hand. A schematic algorithm is given in
\citet{Rasmussen:2005:GPM:1162254}.

\begin{figure*}
\centering
\includegraphics[width=\linewidth]{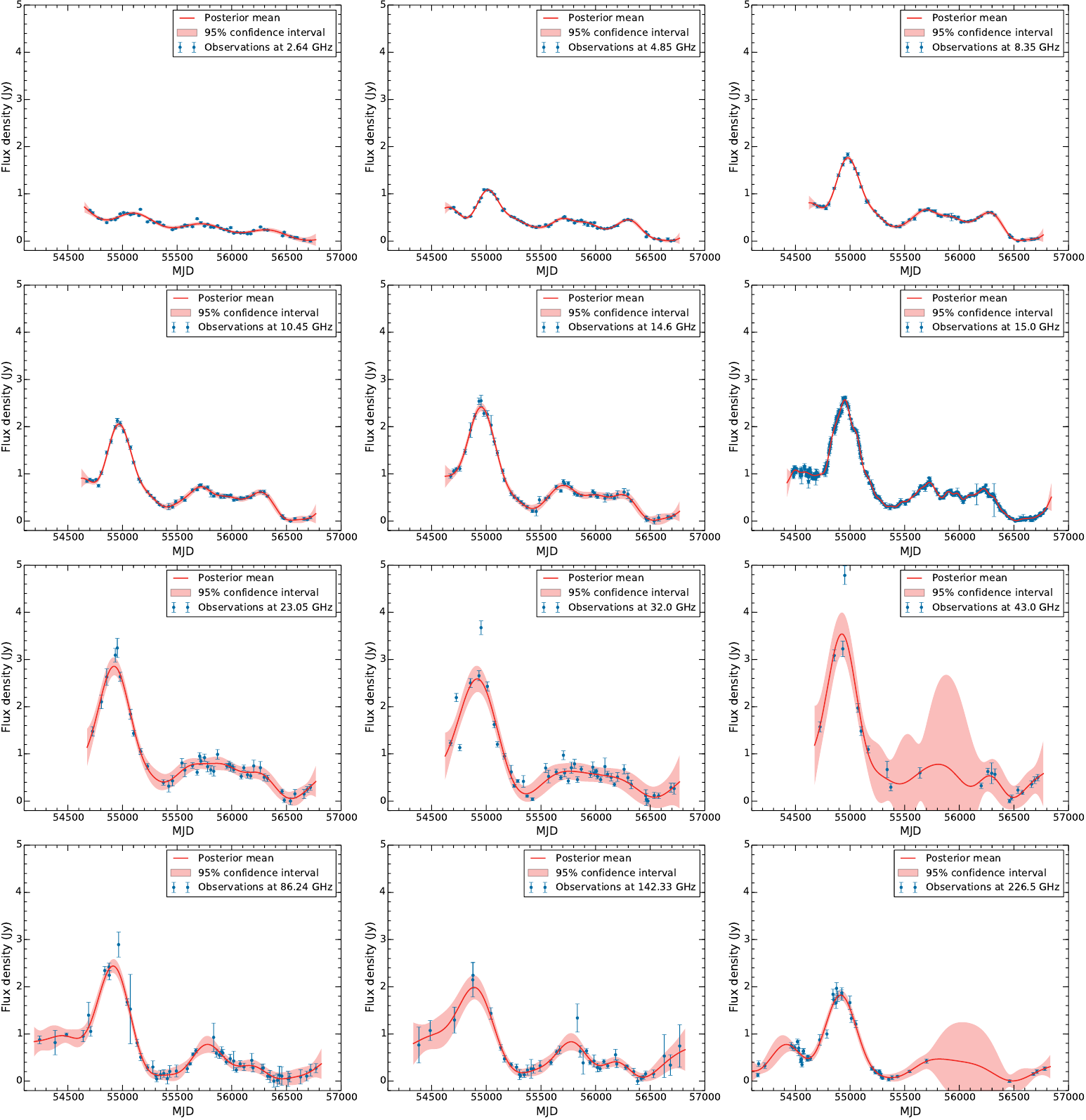}
\caption{Gaussian process regression curves for the radio light curves
in the range 2.64--226.50\,GHz, shown here with the minimum flux density,
$S_{0}$, subtracted. Observations are shown in blue, the posterior mean
(prediction curve) in red, and the 95\% confidence interval is the lighter
red-shaded area.}
\label{fig:1}%
\end{figure*}

For the application of the method to the radio light curves of
PKS\,1502+106, a variant of the algorithm presented in \citet{scikit-learn}
has been used, developed specifically for this purpose. The complete suite of
machine learning programs can be accessed
online\footnote{\url{www.scikit-learn.org/stable/index.html}}.
Here, we consider the full length of each available light curve. Since
consequent flares start without necessarily reaching a quiescent
level prior to the onset of a new one, before proceeding with
the Gaussian process regression, we subtract the minimum flux density level
observed at each frequency, $S_{0}$, from the corresponding light curve. To
ensure the best unbiased result we perform the process with 100 random
initializations of the length scale parameter, $l$. Upon completion, the
posterior mean is returned along with a robust uncertainty estimate in the form
of a 95\% confidence interval for the flux density, and the set of
hyperparameters maximizing the log marginal likelihood. Consequently, we are
able to extract the flare amplitude, $S_{\rm m}$, the time of maximum flux
density, $t_{\rm m}$, and the cross-band delay, $\tau_{\rm GP}$. Additionally,
the flare rising and decay times ($t_{\rm r}$ and $t_{\rm d}$, respectively) are
extracted from the times of the two flux-density minima bracketing $S_{\rm m}$.
In Fig. \ref{fig:1} the results of regression at each frequency are visualized.
The values of $S_{\rm m}$, $t_{\rm m}$, $t_{\rm r}$, and $t_{\rm d}$
characterizing the flare are visible in Tab. \ref{tab:1}.
\begin{table}
    \small
    \centering
    \caption{Results of the Gaussian process regression.}
 \label{tab:1}
 \begin{tabular}{c c c c c c c}
 \hline\hline
    $\nu$ & $S_{0}$ & $S_{\rm m}$ & $t_{\rm m}$ & $t_{\rm r}$ & $t_{\rm d}$ \\
    (GHz) &   (Jy)  &     (Jy)    &     (MJD)   &     (d)     &    (d)      \\
 \hline\noalign{\medskip}
 $  2.64$   &  $0.868$  &  $0.593 \pm 0.032$  & $55095.2$   &  $239.9 $   & $ 366.8 $ \\	       
 $  4.85$   &  $0.777$  &  $1.084 \pm 0.031$  & $55016.8$   &  $201.8 $   & $ 383.2 $ \\	       
 $  8.35$   &  $0.740$  &  $1.764 \pm 0.043$  & $54980.4$   &  $230.9 $   & $ 435.9 $ \\	       
 $ 10.45$   &  $0.725$  &  $2.055 \pm 0.062$  & $54968.7$   &  $234.9 $   & $ 435.0 $ \\	       
 $ 14.60$   &  $0.683$  &  $2.414 \pm 0.088$  & $54954.4$   &  $288.1 $   & $ 467.6 $ \\	       
 $ 15.00$   &  $0.711$  &  $2.570 \pm 0.026$  & $54952.3$   &  $276.2 $   & $ 391.7 $ \\	       
 $ 23.05$   &  $0.478$  &  $2.856 \pm 0.189$  & $54921.9$   &  $390.8 $   & $ 457.0 $ \\	       
 $ 32.00$   &  $0.516$  &  $2.590 \pm 0.277$  & $54916.7$   &  $447.9 $   & $ 446.5 $ \\	       
 $ 43.05$   &  $0.299$  &  $3.542 \pm 0.454$  & $54925.8$   &  $383.4 $   & $ 527.2 $ \\	       
 $ 86.24$   &  $0.507$  &  $2.439 \pm 0.144$  & $54912.8$   &  $331.3 $   & $ 437.2 $ \\	       
 $142.33$   &  $0.263$  &  $1.987 \pm 0.254$  & $54888.0$   &  $313.5 $   & $ 444.5 $ \\	       
 $226.50$   &  $0.238$  &  $1.823 \pm 0.102$  & $54914.3$   &  $300.3 $   & $ 460.8 $ \\	       
 \noalign{\smallskip}
 \hline
 \end{tabular}
 \tablefoot{Columns:
(1) observing frequency,
(2) minimum flux density subtracted prior to the fit,
(3) flare amplitude,
(4) time of flare maximum,
(5) flare rising, and 
(6) decay time.
An uncertainty due to temporal resolution of up to 0.3\,d can be expected
for $t_{\rm m}$, yielding an uncertainty of about 0.4\,d for $t_{\rm r}$ and
$t_{\rm d}$.}
\end{table}

\subsection{The discrete cross-correlation function}

Putative correlated variability across observing bands, based on
sampling-rate-limited time series, can be investigated and
quantified using the discrete cross-correlation function
\citep[DCCF,][]{1988ApJ...333..646E}.

The cross-correlation function (CCF), as a function of the time lag
$\tau$, for two discrete and evenly sampled light curves, $x(t_{i})$ and
$y(t_{i})$, is given by
\begin{equation}
 {\rm CCF}(\tau) = \dfrac{1}{N} \sum_{i=1}^{N} \dfrac{ \left[ x(t_{i})-\bar{x}
\right] \left[ y(t_{i}-\tau)-\bar{y} \right]}{\sigma_{x} \sigma_{y}},
\end{equation}
where
\[
\begin{array}{lp{0.8\linewidth}}
  \bar{x}    & is the mean value of the light curve $x(t_{i})$, \\
  \sigma_{x} & its standard deviation, \\
  \bar{y}    & is the mean of the light curve $y(t_{i})$, and \\
  \sigma_{y} & is the standard deviation of light curve $y(t_{i})$.\\
\end{array}
\]
Uneven sampling of light curves introduces the need for the discrete
cross-correlation function \citep[][see also
\citealp{2012arXiv1207.1459L, 2015MNRAS.453.3455R}]{1988ApJ...333..646E}.
Contrary to linear
interpolation methods, the DCCF takes into consideration only
the data points themselves.
For two irregularly sampled light curves, the unbinned CCF (UCCF) is
\begin{equation}
 {\rm UCCF}_{ij}(\tau) = \dfrac{ \left( x_{i} - \bar{x}
\right) \left(y_{i}-\bar{y} \right) }
{\sqrt{ ( \sigma_{x}^{2} - \bar{\varepsilon}_{x}^{2} ) 
 ( \sigma_{y}^{2} - \bar{\varepsilon}_{y}^{2} ) } },
\end{equation}
where $\bar{\varepsilon}$ is the mean measurement uncertainty of each data
set. For the calculation of the mean and standard deviation, only data within
each time lag bin are considered.
Averaging the UCCF over the number of data pairs, $M$, yields the DCCF
$({\rm DCCF} = \frac{1}{M} \sum {\rm UDCF}_{ij}$) and an uncertainty associated
with each time lag can be determined as
\begin{equation}
 \sigma_{\rm DCCF}(\tau) = \dfrac{1}{M-1} 
 \sqrt{ \sum{ \left[ {\rm UDCF}_{ij} - {\rm DCCF}(\tau) \right] }^{2} }.
\end{equation}
Positive DCCF values indicate a positive correlation
with an average time shift
$\tau$, while negative values imply anti-correlation.

Using the DCCF, \citet{2014MNRAS.441.1899F} searched for possible correlations
between the 3\,mm radio waveband and the \textit{Fermi} $\gamma$-ray
observations for the F-GAMMA blazar sample. Their study has shown the presence
of significantly correlated 3\,mm/$\gamma$-ray variability in an average sense
(whole sample). Specifically, for PKS\,1502{+}106 a 3\,mm/$\gamma$-ray
correlation is detected at a significance level above 99\%, with the mm-radio
emission lagging behind $\gamma$ rays by $14 \pm 11$ days. The aforementioned
difference in the time of arrival translates into a de-projected relative
separation of the corresponding emitting regions of 2.1\,pc, with the
$\gamma$-ray active region upstream of the 3\,mm (or 86.24\,GHz) core. 

\begin{figure}
\centering
\includegraphics[width=\linewidth]{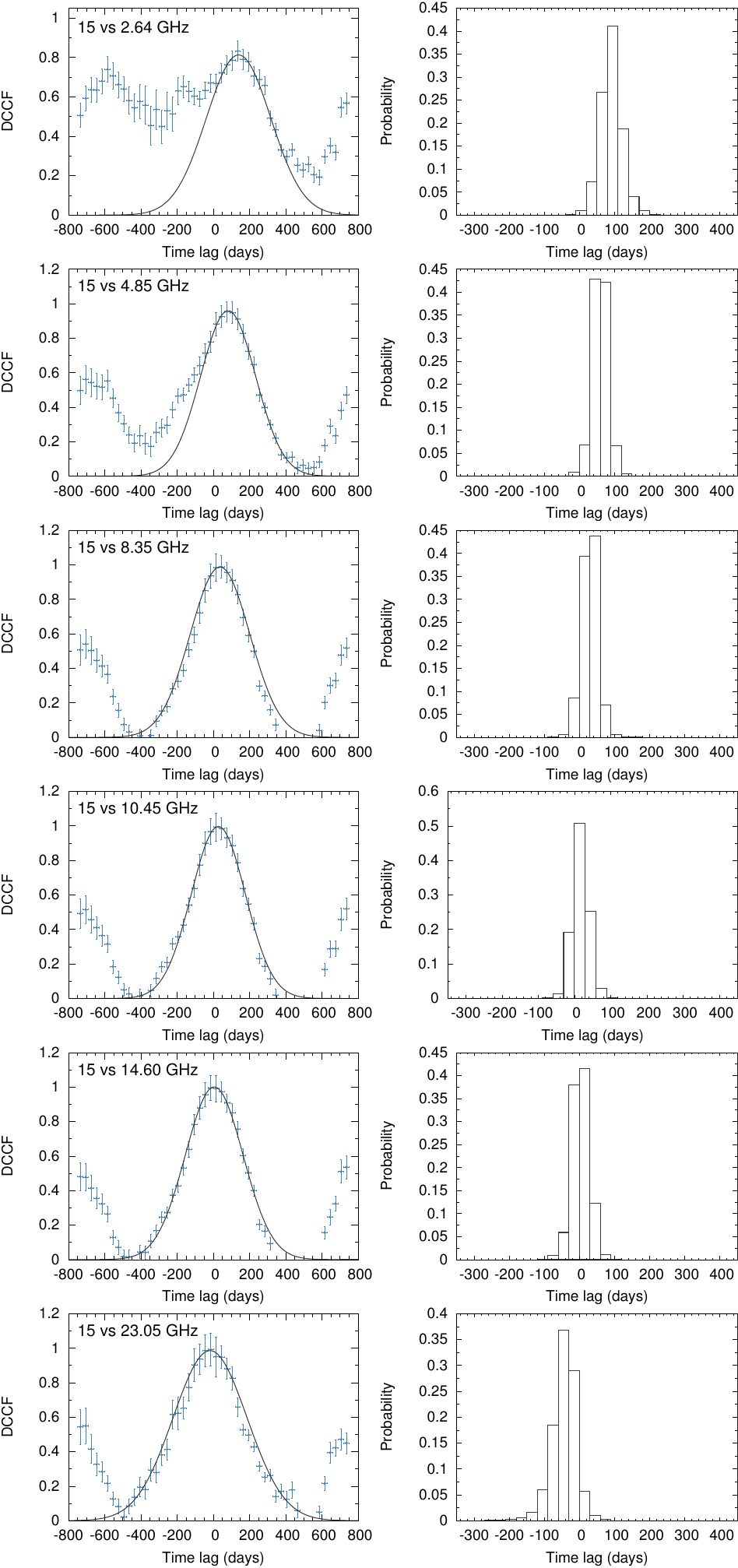}
\caption{DCCFs between the 15\,GHz light curve and those in the range 2.64
to 23.05\,GHz. Solid lines represent the best-fit Gaussian functions.}
\label{fig:2}%
\end{figure}

\begin{figure}
\centering
\includegraphics[width=\linewidth]{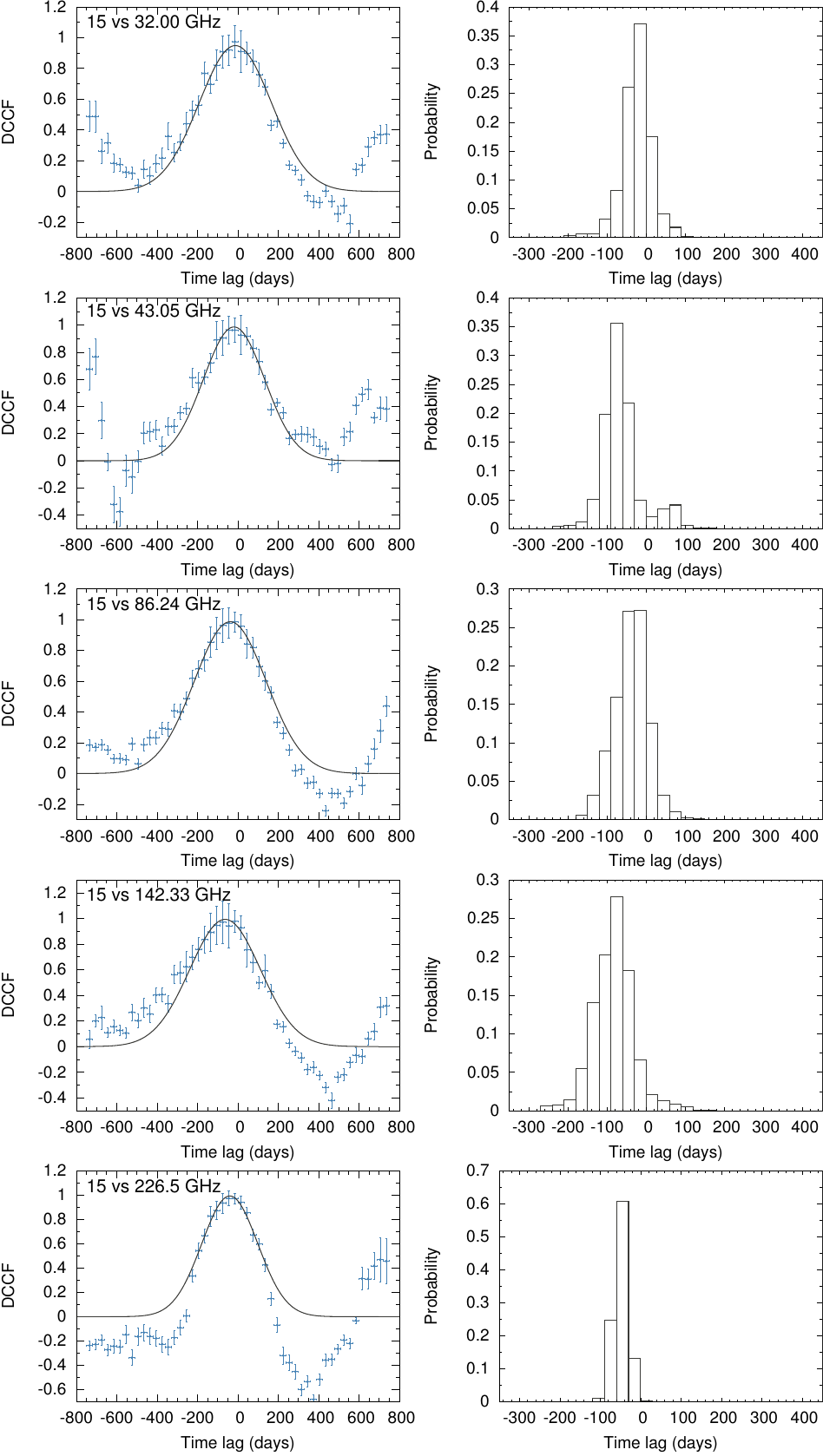}
\caption{DCCFs between the 15\,GHz light curve and those in the range 32.00
to 226.5\,GHz. Solid lines represent the best-fit Gaussians.}
\label{fig:3}%
\end{figure}

Here, we enhance these findings and quantify the correlated variability
beyond the mm/sub-mm band using all available radio observations (see Fig.
\ref{fig:1} for the light curves). The DCCFs have been calculated between the
15\,GHz light curve of the OVRO 40\,m blazar monitoring
program and all available light curves, from 
226.5 down to 2.64\,GHz. For each DCCF we employed a time lag bin width
of 30 days within a window of $\pm 750$\,d. The better sampling of the OVRO
15\,GHz light curve led to its selection as reference. Hereafter, we refer to
the cross-correlated light curves as the reference (OVRO 15\,GHz) and the target
(all the rest).
A positive lag $\tau_{\rm DCCF}$ implies that the target waveband, lags
behind the reference frequency by $\tau_{\rm DCCF}$, measured in days.
We use a weighted Gaussian fit, of the form 
$G(x) = \alpha \, \exp \left[ \frac{-(x-\mu)^{2}}{2\sigma^{2}} \right]$, to
obtain the DCCF peak and its position. Since the value of interest is the peak
of the DCCF, the Gaussian fit is performed in the time lag range that ensures
better peak determination. The mean of the Gaussian is the value of time lag
$\tau_{\rm DCCF}$ reported in Tab. \ref{tab:2}. Left panels of Figs.
\ref{fig:2} and \ref{fig:3} show all DCCFs with respect to our
selected reference frequency along with the Gaussian fit.

For estimating the time-lag uncertainty we employ the model-independent Monte
Carlo (MC) approach introduced by \cite{1998PASP..110..660P} \citep[see
also][]{2003A&A...402..151R}. The two main sources of uncertainty in
cross-correlation analyses -- in addition to limited data trains and
consequently limited number of events -- are: (i) flux uncertainties and (ii)
uncertainties associated with the uneven sampling of the light curves. The
method comprises two parts; that of the flux redistribution (FR), and that of
a random sample selection (RSS), accounting for (i) and (ii) respectively.
The algorithm is essentially as follows:
\begin{enumerate}
 \item Introduction of Gaussian deviates to the data, constrained by the maximum
flux density uncertainty.
 \item Use of a re-sampling with replacement scheme to randomly select data
points and create an equally long light curve; i.e. a bootstrap-like
procedure.
 \item Calculation of the DCCF and determination of the time lag.
 \item After a given number of MC realizations, obtain the
cross-correlation peak distribution (CCPD).
\end{enumerate}

The uncertainty, $\pm \Delta \tau_{68\%}$, or simply $\pm \Delta \tau$, is
obtained directly from the CCPD and corresponds to the time lag interval of 
($\tau_{\rm median} - \Delta \tau$) and ($\tau_{\rm median} + \Delta \tau$),
such that $68\%$ of the realizations yield results within this interval. This
estimate is equivalent to the $1\sigma$ error, in case of a Gaussian
distribution \citep{1998PASP..110..660P}. Here, 5000 MC realizations
were performed.
\begin{table}
    \small
    \centering
    \caption{Results of the correlation analysis at all frequencies.}
 \label{tab:2}
 \begin{tabular}{c c c c}
 \hline\hline
  $\nu$   & $\alpha$ & $\mu$ & $\sigma$ \\
  (GHz)   &          &  (d)  &   (d)    \\
 \hline\noalign{\medskip}
  2.64  & $ 0.81 \pm 0.01 $ &  $139.87 \pm 8.15$  &  $ 178.05 \pm 39.64 $\\ 
  4.85  & $ 0.96 \pm 0.01 $ &  $ 82.12 \pm 1.03$  &  $ 151.60 \pm  2.37 $\\ 
  8.35  & $ 0.99 \pm 0.01 $ &  $ 37.19 \pm 1.98$  &  $ 164.54 \pm  4.81 $\\ 
 10.45  & $ 0.99 \pm 0.01 $ &  $ 28.11 \pm 2.85$  &  $ 149.57 \pm  7.61 $\\ 
 14.60  & $ 1.00 \pm 0.01 $ &  $  5.15 \pm 1.05$  &  $ 159.42 \pm  3.94 $\\ 
 23.05  & $ 0.98 \pm 0.01 $ &  $-17.74 \pm 3.48$  &  $ 204.18 \pm 10.01 $\\ 
 32.00  & $ 0.95 \pm 0.01 $ &  $-12.31 \pm 3.24$  &  $ 179.45 \pm  5.99 $\\ 
 43.05  & $ 0.99 \pm 0.01 $ &  $-19.04 \pm 5.38$  &  $ 152.45 \pm  7.03 $\\ 
 86.24  & $ 0.98 \pm 0.01 $ &  $-37.67 \pm 4.34$  &  $ 178.78 \pm 10.90 $\\ 
142.33  & $ 0.99 \pm 0.02 $ &  $-62.98 \pm 5.88$  &  $ 175.25 \pm 15.00 $\\ 
226.50  & $ 0.99 \pm 0.02 $ &  $-39.89 \pm 4.84$  &  $ 138.46 \pm  9.00 $\\
\noalign{\smallskip}
 \hline
 \end{tabular}
 \tablefoot{Columns:
 (1) observing frequency,
 (2) DCCF peak value,
 (3) time lag value of the peak, and
 (4) width of the Gaussian function.}
\end{table}

%
\section{Results}\label{sect:results}

\subsection{Flare parameters}

\begin{figure}
\centering
\includegraphics[width=\linewidth]{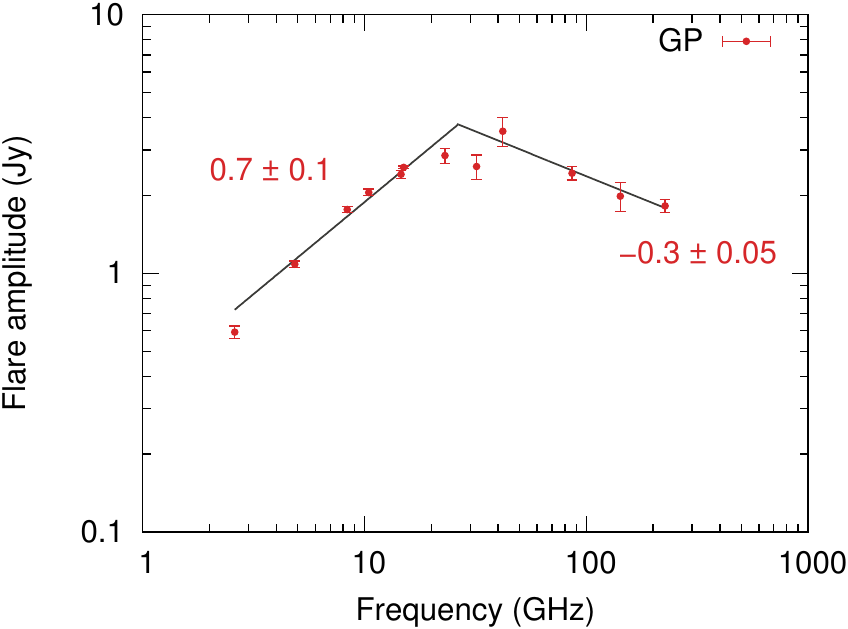}
\caption{Frequency dependence of the flare amplitudes. Solid lines show the
wings of the fitted broken power law, while the values denote their slopes.}
\label{fig:4}%
\end{figure}

The flare amplitudes, $S_{\rm m}$, show an increasing trend with frequency, up
to about 43\,GHz whereafter they start decaying again as seen for 86.24,
142.33, and 226.5\,GHz (see Fig. \ref{fig:4}). As will be
discussed later, there exists a clear tendency for the flare to
be visible earlier at higher frequencies, an effect attributed to synchrotron
opacity (see also Fig. \ref{fig:5}). The flare rising times, while initially
increase with frequency, at frequencies higher than 32.00\,GHz show a
decreasing trend (see Tab. \ref{tab:1}). Finally, in terms of the flare decay
times they seem to be following a slightly increasing trend with observing
frequency in the whole range employed here (Tab. \ref{tab:1}).

\subsection{Frequency-dependent time lags}

The DCCFs shown in Figs. \ref{fig:2} and \ref{fig:3} confirm the
presence of a correlation between the outburst seen across observing
frequencies. Essentially in all cases, a single strong peak is clearly
visible. In the following, we use the obtained time lag where the DCCF peaks
(Tab. \ref{tab:2}), referenced to the highest-frequency, well-sampled light
curve at 142.33\,GHz. The final $\tau_{\rm DCCF}$ are reported in Tab.
\ref{tab:3}. 

\begin{figure}
\centering
\includegraphics[width=\linewidth]{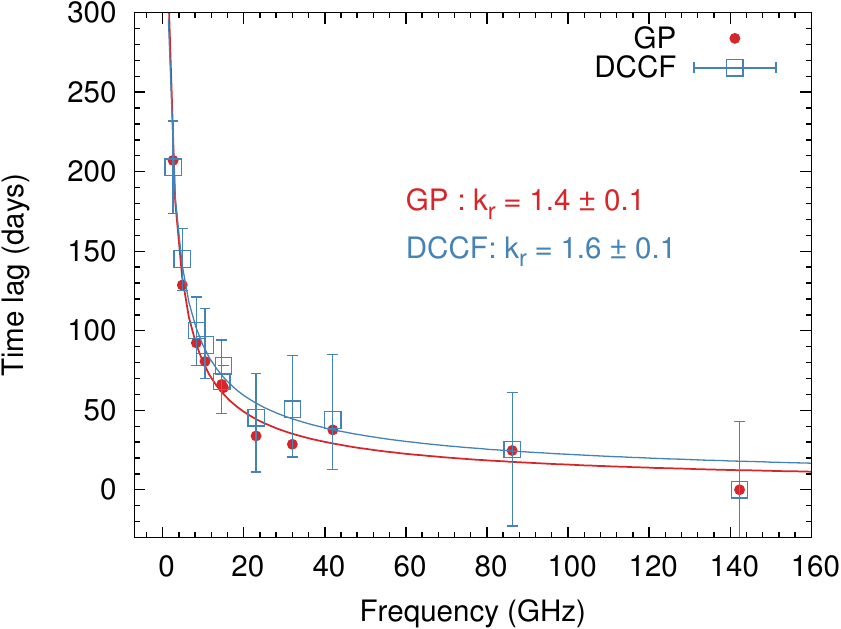}
\caption{Frequency dependence of the time lags with respect to the data at
142.33\,GHz. Red circles denote values resulting from the GP regression and
blue squares those from the DCCF analysis. Solid lines represent the best-fit
power laws.}
\label{fig:5}%
\end{figure}

In Fig. \ref{fig:5} the delays obtained with both methods are shown. The
GP regression and the DCCF analysis yield conclusive results that agree very
well. A systematic trend is clearly seen, in the sense that the time lags
become larger towards lower frequencies. This trend is described well by a
power-law. In order to obtain its index, the frequency-dependent delays are
fitted by a power-law of the form $a \nu^{-1/k_{\rm r}}$ using a least-squares
procedure. The results for index $k_{\rm r}$ are for the GP
regression $k_{\rm r, \,GP} = (1.4 \pm 0.1)$ 
and $k_{\rm r, \,DCCF} = (1.6 \pm 0.1)$
for the DCCF results. Both are shown, along with the best-fit curves, in
Fig. \ref{fig:5}.

The rather conservative uncertainty estimates of the FR/RSS method
\citep{1998PASP..110..660P} and the good agreement between $\tau_{\rm DCCF}$ and
$\tau_{\rm GP}$ allow an average time lag, $\left< \tau \right>$, and its
standard error to be obtained (Tab. \ref{tab:3}). Hereafter, this average
value is used for all calculations and all uncertainties are formally
propagated. The longest delay of $(205 \pm 2.0)$\,days is seen between
$142.33$ and $2.64$\,GHz. A decreasing trend is clearly visible towards higher
frequencies with a time lag of $(136.9 \pm 8.1)$\,d at 4.85\,GHz and
$\left< \tau \right> = (96.2 \pm 3.8)$\,d at 8.35\,GHz. Between the two
highest frequencies of 142.33\,GHz and 86.24\,GHz, it is $\left< \tau
\right> = (24.8 \pm 0.2)$\,days (see Tab. \ref{tab:3}).
%
\begin{table*}
    \small
    \centering
    \caption{Frequency-dependent time lags and ``time-lag core shifts'' with
respect to the observations at $142.33$\,GHz.}
 \label{tab:3}
 \begin{tabular}{c c c c c c c c c}
 \hline\hline
$\nu$ & $\tau_{\rm GP}$ & $\tau_{\rm DCCF}$ & $\left< \tau \right>$ & $\Delta r$ & $\Omega_{{\rm r} \nu}$ \\
(GHz) & 	(d)	&	    (d)     &	  (d)		    &	  (mas)  &	  (pc\,GHz)	  \\
 \hline\noalign{\medskip}
 \smallskip
 $  2.64$   &  $207.1$   &  $203 \, ^{+29}  _{-29}$  &     $205.0 \pm 2.0 $   &  $ 0.051 \pm  0.005  $    &   $ 0.885 \pm 0.093 $    \\ 	
 \smallskip 		
 $  4.85$   &  $128.8$   &  $145 \, ^{+19}  _{-20}$  &     $136.9 \pm 8.1 $   &  $ 0.034 \pm  0.004  $    &   $ 0.922 \pm 0.118 $    \\ 	
 \smallskip 		
 $  8.35$   &  $92.4 $   &  $100 \, ^{+21}  _{-22}$  &     $96.2  \pm 3.8 $   &  $ 0.024 \pm  0.003  $    &   $ 0.981 \pm 0.127 $    \\ 	
 \smallskip 		
 $ 10.45$   &  $80.6 $   &  $91  \, ^{+23}  _{-21}$  &     $85.8  \pm 5.2 $   &  $ 0.021 \pm  0.002  $    &   $ 1.046 \pm 0.148 $    \\ 	
 \smallskip 		
 $ 14.60$   &  $66.3 $   &  $68  \, ^{+26}  _{-20}$  &     $67.2  \pm 0.9 $   &  $ 0.017 \pm  0.002  $    &   $ 1.081 \pm 0.147 $    \\ 	
 \smallskip 		
 $ 15.00$   &  $64.2 $   &  $63  \, ^{+43}  _{-51}$  &     $63.6  \pm 0.6 $   &  $ 0.016 \pm  0.002  $    &   $ 1.047 \pm 0.142 $    \\ 	
 \smallskip 		
 $ 23.05$   &  $33.8 $   &  $45  \, ^{+28}  _{-34}$  &     $39.4  \pm 5.6 $   &  $ 0.010 \pm  0.002  $    &   $ 0.955 \pm 0.194 $    \\ 	
 \smallskip
 $ 32.00$   &  $28.6 $   &  $51  \, ^{+34}  _{-30}$  &     $39.8  \pm 11.2$   &  $ 0.010 \pm  0.003  $    &   $ 1.338 \pm 0.428 $    \\ 	
 \smallskip
 $ 43.05$   &  $37.7 $   &  $44  \, ^{+41}  _{-31}$  &     $40.8  \pm 3.1 $   &  $ 0.010 \pm  0.001  $    &   $ 1.915 \pm 0.339 $    \\ 	
 \smallskip
 $ 86.24$   &  $24.7 $   &  $25  \, ^{+36}  _{-48}$  &     $24.8  \pm 0.2 $   &  $ 0.006 \pm  0.001  $    &   $ 3.585 \pm 0.623 $    \\ 	
  \noalign{\smallskip}
 \hline
 \end{tabular}
 \tablefoot{Columns:
(1) observing frequency,
(2) time lag from GP regression,
(3) time lag from DCCF,
(4) average of time lags $\tau_{\rm GP}$ and $\tau_{\rm DCCF}$,
(5) the angular offset in units of mas [$\Delta r = \mu \left< \tau \right>$],
and
(6) time-lag-based core position offset.
An uncertainty of up to 0.4\,d can be associated with $\tau_{\rm GP}$.}
\end{table*}

\subsection{Time-lag core shifts and synchrotron opacity structure} 

Like most blazars, PKS\,1502+106 features a flat-spectrum core, indicative of
the superposition of  optically thick synchrotron components
\citep[e.g.][]{2006MNRAS.367.1083K}. In addition to the scenario that the
observed VLBI core is the first stationary hotspot (recollimation shock) of the
jet \citep{1988ApJ...334..539D, 1995ApJ...449L..19G}, a widely accepted
interpretation is that this core is the point of the continuous flow where the
photons at a given frequency have sufficient energy to escape the
opacity barrier. Within this scenario, the core is the surface
where the optical depth is close to unity at a given frequency
\citep[e.g.][]{1981ApJ...243..700K}. The standard relativistic jet
paradigm \citep{1979ApJ...232...34B,1981ApJ...243..700K} predicts that the
apparent position of this unit-opacity surface depends on observing frequency.
Therefore, according to the interpretation of the VLBI core discussed above,
its position must also be frequency-dependent. This ``core-shift effect'' was
first observed by \cite{1984ApJ...276...56M} and ever since has been studied
using multi-frequency VLBI
\citep[e.g.][]{1998A&A...330...79L, 2008A&A...483..759K, 2011A&A...535A..24S,
2012A&A...545A.113P}.
The importance of core-shift measurements lies in the fact that they can
provide critical insights into the structure and physics of ultracompact
jets, such as the position of the core at each frequency relative 
to the jet base, and the magnetic field in different locations along the
relativistic flow \citep[e.g.][]{1998A&A...330...79L, 2005ApJ...619...73H}.

Relative timing of total flux density outbursts, as seen in single-dish radio
observations, presents a viable alternative to VLBI core shifts. Let us return
back to the standard jet model. Here, a disturbance
originating at the jet base will propagate outwards along the jet. As this
disturbance or shock moves away from the base it crosses the sequence of
physically separated regions where the jet flow becomes optically thin -- i.e.
the VLBI cores -- as seen at a given sequence of observing frequencies. When the
disturbance reaches a given unit-opacity surface for a given frequency $\nu_{\rm
i}$, the synchrotron flare emission most likely becomes optically thin at
$\nu_{\rm i}$. This appears in single-dish light curves as
a total intensity outburst, while in VLBI maps these travelling shocks manifest
as regions of enhanced emission after the core, the so-called VLBI
knots \citep{1985ApJ...298..114M}. Since the times of maxima in
single-dish light curves correspond to the flare emission becoming optically
thin at the unit-opacity surface, for a given frequency, it follows that these
times are frequency-dependent as well, and carry information on the opacity of
each core. In other words, through relative timing of the same flaring
event seen at different bands, we can obtain a tomography of the jet in terms
of nuclear opacity, similar to VLBI core-shift measurements
\citep[e.g.][]{2011MNRAS.415.1631K, 2014MNRAS.437.3396K}.

An important consideration to be taken into account is that the method is
based on the absence of any optically thick component downstream of the
core. If there was any, it would sooner or later induce an
opacity-driven enhancement of flux density due to its expansion and subsequent
drop of density at a region further away from the core, the position of which we
try to constrain. Essentially, the emission needs to become optically thin when
it reaches each respective unit-opacity surface. In this respect and given the
spectra of VLBI components presented in \cite{2016A&A...586A..60K}, component
C3, which is used for our calculations, is optically thin at
least between 43 and 86\,GHz (spectral index $\alpha_{\rm 43/86\,GHz} =
-0.8 \pm 0.3$).

Based on \cite{1998A&A...330...79L}, a ``time-lag core-shift''
between two frequencies $\nu_{1}$ and $\nu_{2}$, with $\nu_{2}$ > $\nu_{1}$, can
be defined as follows \cite[see also][]{2011MNRAS.415.1631K}
\begin{equation}
 \Omega_{{\rm r} \nu} = 4.85 \times 10^{-9}
 \dfrac{\mu \, \Delta t \, \rm{D_{L}}}{(1+z)^{2}}
 \left(
 \dfrac{\nu_{1}^{1/k_{\rm r}} \cdot \nu_{2}^{1/k_{\rm r}}}{\nu_{2}^{1/k_{\rm r}}
- \nu_{1}^{1/k_{\rm r}}}
 \right), 
 \label{eq:omega}
\end{equation}
in units of pc\,GHz for $k_{r}=1$, where
\[
\begin{array}{lp{0.8\linewidth}}
  \mu        & is the VLBI jet proper motion in mas\,yr$^{-1}$,\\
  \Delta t   & is the observed time lag between the frequency pair $\nu_{1}$
                and $\nu_{2}$ in yr, \\
  \rm{D_{L}} & is the luminosity distance in pc, \\
  z          & is the redshift,\\
\nu_{1}, \nu_{2} & are the first and second observing frequency,
                  both expressed in GHz, and \\
k_{\rm r}     & is the power-law index obtained by fitting the cross-band
delays.
\end{array}
\]
Here, $k_{\rm r} = [(3-2 \alpha)m + 2n -2]/(5 - 2 \alpha)$ with $m$ and $n$
denoting the index of the power-law dependence of the magnetic field $B(r)
\propto r^{-m}$ and the electron number density $N_{\rm e}(r) \propto r^{-n}$,
at distance $r$ along the jet, respectively. Finally, $\alpha$ denotes the
optically thin spectral index. In case of external density gradients and/or
free-free absorption, it is $k_{\rm r} > 1$. For synchrotron self-absorption
(SSA) dominated opacity and equipartition between particle energy and magnetic
field energy, $k_{\rm r} = 1$ for $m = 1$ and $n = 2$ independently of $\alpha$.
The angular offset in units of mas is given by $\Delta r = \mu \left< \tau
\right>$. That is, the distance traveled by the disturbance within the time
interval $\left< \tau \right>$ (Tab. \ref{tab:3}).
Using $\Omega_{{\rm r} \nu}$ and  following \cite{1998A&A...330...79L},
\cite{2005ApJ...619...73H}, and \cite{2011MNRAS.415.1631K} we obtain
the distance between the core and the base of the assumed-conical jet. The
frequency-dependent distance of the core -- at any frequency $\nu$ -- is
given by
\begin{equation}
 r_{\rm core} (\nu) = \dfrac{\Omega_{{\rm r} \nu}}{\sin \theta} \nu^{-1/k_{\rm
r}}.
\label{eq:rcore}
\end{equation}

We employ Eqs. \ref{eq:omega} and \ref{eq:rcore} to infer the separation of the
radio core at each frequency from the jet base. In that, we use the average
$k_{\rm r} = (1.5 \pm 0.1)$ calculated based on the DCCF
analysis and the Gaussian process regression. Both agree very well with each
other indicating that $k_{\rm r} > 1$. This implies the
possible presence of jet external density gradients and/or foreground free-free
absorption. Other parameters used are $\mu =0.09$\,mas\,yr$^{-1}$ which
corresponds the proper motion of component C3 at 86\,GHz, responsible for the
flare, and $\theta = 2.6^{\circ}$ the critical viewing angle
\citep{2016A&A...586A..60K}.

Our results reveal the nuclear opacity structure of PKS\,1502+106 (see Tab.
\ref{tab:4}). The largest distance of about 10\,pc is seen for the 2.64\,GHz
core, as expected, and distances diminish as we go towards higher frequencies.
At observing frequencies between 23.05 and 43.05\,GHz the light curves become
slightly more sparsely sampled due to the fact that weather effects and poorer
system performance of the Effelsberg 100-m telescope, at those bands, play an
increasingly important role. 
At the highest frequency of 86.24\,GHz, the core lies at a distance of
$(4.0 \pm 1.1)$\,pc away from the jet base.

\begin{table*}
    \centering
    \caption{Synchrotron opacity structure and magnetic field tomography of
the jet of PKS\,1502+106.}
 \label{tab:4}
 \begin{tabular}{c c c c}
 \hline\hline
$\nu$  &    $r_{\rm core}$   &  $B_{1\,{\rm pc}} ~~ [B_{1\,{\rm pc\,min}}, B_{1\,{\rm pc\,max}}]$   & $B_{\rm core} ~~ [B_{\rm core\,min}, B_{\rm core\,max}]$ \\ 
(GHz)  &       (pc)	     &  	(mG)	         &	 (mG)	   \\
 \hline\noalign{\medskip}
  2.64 &  $ 10.2 \pm 1.2$  &  $ 148 ~~~~ [106, 209]$   &   $ ~~ 14 ~~~~~~ [ 11,  20] $  \\ 
  4.85 &  $  7.1 \pm 1.0$  &  $ 154 ~~~~ [111, 220]$   &   $ ~~ 22 ~~~~~~ [ 16,  30] $  \\	
  8.35 &  $  5.3 \pm 0.8$  &  $ 166 ~~~~ [121, 233]$   &   $ ~~ 32 ~~~~~~ [ 24,  43] $  \\	
 10.45 &  $  4.8 \pm 0.8$  &  $ 178 ~~~~ [129, 253]$   &   $ ~~ 37 ~~~~~~ [ 27,  51] $  \\	
 14.60 &  $  4.0 \pm 0.7$  &  $ 185 ~~~~ [137, 259]$   &   $ ~~ 46 ~~~~~~ [ 35,  63] $  \\	
 15.00 &  $  3.8 \pm 0.7$  &  $ 178 ~~~~ [132, 249]$   &   $ ~~ 47 ~~~~~~ [ 36,  64] $  \\	
 23.05 &  $  2.6 \pm 0.6$  &  $ 161 ~~~~ [113, 237]$   &   $ ~~ 62 ~~~~~~ [ 45,  88] $  \\	
 32.00 &  $  2.9 \pm 1.0$  &  $ 235 ~~~~ [142, 398]$   &   $ ~~ 80 ~~~~	[ 51, 131] $  \\	
 43.05 &  $  3.4 \pm 0.8$  &  $ 352 ~~~~ [249, 522]$   &   $   102 ~~~~	[ 76, 145] $  \\	
 86.24 &  $  4.0 \pm 1.1$  &  $ 711 ~~  [494, 1086]$   &   $   176 ~~	[129, 254] $  \\     
  \noalign{\smallskip}
 \hline
 \end{tabular}
 \tablefoot{Columns:
(1) observing frequency,
(2) distance of the radio core to the jet base,
(3) magnetic field strength at a distance of 1\,pc from the jet base, and
(4) core magnetic field strength. In cols. 3 and 4, bracketed values correspond
to the minimum and maximum values of the plausible range, accounting for
the uncertainty of the inputs.}
\end{table*}

\subsection{Equipartition magnetic field}\label{sect:Bfield}

We obtain an expression for the magnetic field at 1\,pc from the jet base,
$B_{1\,{\rm pc}}$, by evaluating Eq. (43) of \cite{2005ApJ...619...73H}
\citep[see also][]{2009MNRAS.400...26O, 2014MNRAS.437.3396K}. 
Under equipartition between the energies of the magnetic field and that of
radiating particles, and with a spectral index $\alpha = -0.5$, this yields
\begin{equation}
 B_{1\,{\rm pc}} \approx 0.014 ~ \left[ \dfrac{\Omega_{{\rm r} \nu}^{3k_{\rm r}}
\,(1+z)^{2} \ln(\gamma_{\rm max}/\gamma_{\rm min})}{\delta^{2} \phi \sin^{3
k_{\rm r} - 1}\theta} \right]^{1/4}  ~~~~~ ({\rm G}),
\label{eq:B1pc}
\end{equation}
where
\[
\begin{array}{lp{0.75\linewidth}}
  \gamma_{\rm max}, \gamma_{\rm min} & are the maximum and minimum Lorentz
factors of the emitting $e^{-}$,\\
  \delta & is Doppler factor,\\
  \phi   & is the half-opening angle of the jet, and\\
  \theta & is the viewing angle.
\end{array}
\]
Then, the magnetic field strength at the core is given by
\begin{equation}
 B_{\rm core} (\nu) = B_{1\,{\rm pc}} \, r_{\rm core}(\nu)^{-1}.
 \label{eq:Bcore}
\end{equation}

We adopt $\ln(\gamma_{\rm max}/\gamma_{\rm min}) = 10$ and employ the values
for the Doppler factor and half-opening angle, deduced
in \cite{2016A&A...586A..60K}, of $\delta = 10$ and 
$\phi = (1.90 \pm 0.25)^{\circ}$, respectively corresponding to components
travelling within the inner jet of PKS\,1502+106. Resulting values for
$B_{1\,{\rm pc}}$ and $B_{\rm core}$ are summarized in Tab. \ref{tab:4}. Here,
we also report a conservative range for these values, by combining the
uncertainties of the input parameters in a way that maximizes this range. 
Inferred magnetic field strengths are higher at higher frequencies, when
approaching the jet base. The values range between 14 and 176\,mG for the
2.64\,GHz (outermost) and 86.24\,GHz (innermost) cores, respectively.

The results presented here are in reasonable agreement with the core shift
measurements of PKS\,1502+106 derived by \cite{2012A&A...545A.113P} who employ
standard VLBI techniques and obtain a separation of ${\sim} 8$\,pc between the
15\,GHz VLBI core and the vertex of the jet. Some discrepancies can be seen in
the -- inferred from VLBI core shifts -- physical parameters of the jet, such as
the magnetic field at a distance of 1\,pc and in the 15\,GHz core ($B_{1\,{\rm
pc}}$ and $B_{\rm core}$), but these can be explained through the different
kinematical parameters used. Specifically, \cite{2012A&A...545A.113P} make use
of the fastest non-accelerating, component's apparent speed from data
covering the period 1994 August to 2007 September, as reported in
\cite{2009AJ....138.1874L}. However, here we use the proper
motion, $\mu$, of the knot most likely responsible for the multi-wavelength
flare of 2008/2010 \citep[knot C3 in][]{2016A&A...586A..60K} whose apparent
angular velocity -- while travelling in the inner jet of PKS\,1502+106 -- is
about half of what \cite{2012A&A...545A.113P} use.

%
\section{Discussion}\label{sect:Disc}

\subsection{Localizing the $\gamma$-ray emission}

Our ultimate goal of pinpointing the $\gamma$-ray emission region is reached by
combining the findings of the present study with those
of \cite{2014MNRAS.441.1899F}, who deduce a relative distance between the
$\gamma$-ray active region and the 86.24\,GHz unit-opacity surface of
about 2.1\,pc \citep[cf.][]{2015MNRAS.452.1280R}.
Here, based on the opacity-driven time lags across the eleven observing
frequencies, the distance of the 86.24\,GHz core from the jet base is
$(4.0 \pm 1.1)$\,pc. From the combination follows that the $\gamma$-ray
emission region is best constrained to a distance of $(1.9 \pm 1.1)$\,pc
away from the jet base.
In a similar study, \cite{2014MNRAS.445..428M} obtain a time lag of $(-40 \pm
13)$\,d between the leading $\gamma$-rays and the lagging 15\,GHz radio
emission. Following their argumentation, this time lag translates into a
relative distance of $(2 \pm 1)$\,pc between the 15\,GHz unit-opacity surface
and the $\gamma$-ray site. Combining our $r_{\rm core}$ at 15\,GHz
(Tab. \ref{tab:4}) with the aforementioned value, we obtain a separation of
$(1.8 \pm 1.2)$\,pc between the jet base and the $\gamma$-ray active region,
comparable with our previous result based on the 86.24\,GHz time lag. We
reiterate that in the present work we use the VLBI kinematical parameters of the
knot connected to the 2008/10 flare of PKS\,1502+106 (C3, see Sect.
\ref{sect:Bfield}). Finally, our results are in agreement with the upper limit
of ${\leq} 5.9$\,pc reported by \cite{2016A&A...586A..60K}, based on mm-VLBI
imaging.

At $(1.9 \pm 1.1)$\,pc from the jet base, the high-energy emission
originates well beyond the bulk of the BLR material of PKS\,1502+106, placed at
${\rm R}_{\rm BLR} \approx 0.1$\,pc \citep{2010ApJ...710..810A}. A $\gamma$-ray
emission region placed so far away from the BLR almost negates the notion that
BLR photons alone can be used as the target photon field for inverse Compton
up-scattering to $\gamma$-rays, thus other or additional seed photon fields
(e.g. torus, synchrotron self-Compton, SSC) need to be involved in the process.

\subsection{A shock origin for the 2008/10 flare}

Overall, the slopes extracted from the multi-wavelength light curves agree well
with the expectations of the shock-in-jet model \citep{1985ApJ...298..114M},
indicating a shock origin for the 2008/2010 flare of PKS\,1502+106.
In the following, the observed light curve
parameters (flare amplitudes and cross-band delays) are discussed in
conjunction with the results of analytical shock-in-jet model simulations
\citep{2015A&A...580A..94F}.
The frequency dependence of flare amplitudes is shown in Fig. \ref{fig:4} and
of the cross-frequency delays in Fig. \ref{fig:5}.
Flare amplitudes rise with frequency until they culminate at a frequency
of about 43\,GHz, after which the amplitude of the flare drops. The rise and
decay wings follow a trend that is described well by a broken power law, in very
good agreement with the shock model expectations. The slopes are
$\epsilon_{\rm flare\,amp\,rise} = (0.7 \pm 0.1)$ for the rising part of the
flare amplitude and $\epsilon_{\rm flare\,amp\,decay} = (-0.3 \pm 0.05)$ for the
decaying part. The power-law slope of the observed time lags,
$\epsilon_{\rm time-lag} = -1/k_{\rm r}$ (see Sect. \ref{sect:results}) is
$\epsilon_{\rm time-lag} \approx -0.7$ (see Fig. \ref{fig:5}). This value is
obtained by averaging over the two values arising from the GP regression and the
DCCF analysis and is in accord with the pattern expected from the shock model.
By comparing the slope of the frequency-dependent cross-band delays with the
finding of \cite{2015A&A...580A..94F}, we can safely exclude the scenario of a
decelerating jet. Our findings in fact suggest the presence of a certain degree
of acceleration taking place within the relativistic jet of PKS\,1502+106.
%

%
\section{Conclusions}\label{sect:Conc}

In the present work we exploited the dense, long-term F-GAMMA radio light
curves in the frequency range 2.64--142.33\,GHz obtained with the Effelsberg
100-m and IRAM 30-m telescopes. The F-GAMMA data were complemented by light
curves at 15.00\,GHz (OVRO) and 226.5\,GHz (SMA).

A detailed section was devoted to the two independent approaches used in
order to quantify the observed flare of PKS\,1502+106 in the time domain.
These are: (i) a Gaussian process regression and (ii) a discrete
cross-correlation function analysis.
This is the first time that GP regression is used in the field of blazar
variability and it has shown to perform well. It is a viable approach to the
problem of modelling discrete, unevenly sampled, blazar light curves and
extracting the relevant parameters.

The flare in PKS\,1502+106 is a ``clean'', isolated outburst seen across
observing frequencies. Our findings suggest a frequency-dependence for the
cross-band delays and the flare amplitudes. The flare amplitudes first increase
-- up to the frequency of ${\sim}43$\,GHz, where they peak -- after which a
decreasing trend follows (Fig. \ref{fig:4}). The times at which the flare peaks
are characterized by increasing time delay towards lower frequencies (Fig.
\ref{fig:5}). The characteristic dependence of these parameters on
the observing frequency can be approximated by power laws, indicative of shock
evolution and pc-scale jet acceleration.

Through the observed opacity-driven time lags, the structure of PKS\,1502+106
in terms of synchrotron opacity is deduced -- i.e. using a
``time-lag core shift'' method. The positions of the ten radio unit-opacity
surfaces with respect to the jet base were deduced, with distances in the
range $(10.2 \pm 1.2)$\,pc to $(4.1 \pm 1.1)$\,pc for the core between
2.64 and 86.24\,GHz, respectively.

The frequency-dependent core positions also enable a tomography of
the magnetic field along the jet axis.
We obtain the equipartition magnetic field at the position of each core, $B_{\rm
core}$, and also at a distance of 1\,pc from the jet base, $B_{1\,{\rm pc}}$. 
The former, between the 2.64 and 86.24\,GHz unit-opacity surfaces, are in
the range $B_{\rm core,\,2.64} \sim 14$\,mG and $B_{\rm core,\,86.24} \sim
176$\,mG. For the latter, our figures indicate an average value of $\left<
B_{1\,{\rm pc}} \right> = (174 \pm 55)$\,mG.

The $\gamma$-ray emission region is constrained to $(1.9 \pm 1.1)$\,pc away
from the jet base, well beyond the bulk of the BLR material of
PKS\,1502+106. This yields a contribution of IR torus photons and/or SSC to the
production of $\gamma$ rays in PKS\,1502+106, while almost negates the
contribution of accretion disk or BLR photons alone.

The flare complies with the typical shock evolutionary path and its overall
behavior in the time domain is in accord with the shock-in-jet scenario.
When seen in the light of the VLBI findings discussed in
\cite{2016A&A...586A..60K}, our results corroborate the
scenario that the flare of PKS\,1502+106 was induced by a shock,
seen in mm-VLBI images as component C3, travelling downstream of the
core at 43/86\,GHz maps. This travelling disturbance is associated with the
multi-frequency flare seen from radio up to $\gamma$-ray energies.

\begin{acknowledgements}

V.K. is thankful to B.~Boccardi, A.~P.~Lobanov, W.~Max-Moerbeck, B.~Rani,
S.~Larsson, and A.~Karastergiou for informative and fruitful discussions. We
thank the anonymous referee for useful suggestions.



V.K., I.N., and I.M. were supported for this research through a stipend from the
International Max Planck Research School (IMPRS) for Astronomy and Astrophysics
at the Universities of Bonn and Cologne.

Based on observations with the 100-m telescope of the MPIfR at Effelsberg and
with the IRAM 30-m telescope.

The single-dish mm observations were coordinated with the flux density
monitoring conducted by IRAM, and data from both programs are included here.

IRAM is supported by INSU/CNRS (France), MPG (Germany) and IGN (Spain).

The OVRO 40-m Telescope \textit{Fermi} Blazar Monitoring Program is supported by
NASA under awards NNX08AW31G and NNX11A043G, and by the NSF under awards
AST-0808050 and AST-1109911.

The Submillimeter Array is a joint project between the Smithsonian Astrophysical
Observatory and the Academia Sinica Institute of Astronomy and Astrophysics and
is funded by the Smithsonian Institution and the Academia Sinica.

This research has made use of NASA's Astrophysics Data System.
\end{acknowledgements}

 \bibliographystyle{aa} 
 \bibliography{refs_all}


\end{document}